\documentclass{PoS}

\usepackage{booktabs}
\usepackage{mathrsfs,latexsym}

\newcommand{\eq}[1]{Eq.~(\ref{#1})}
\newcommand{\fig}[1]{Fig.~\ref{#1}}
\newcommand{\tab}[1]{Table~\ref{#1}}
\newcommand{\sect}[1]{Section~\ref{#1}}
\newcommand{\tmtextbf}[1]{{\bfseries{#1}}}

\newcommand{\tmtexttt}[1]{{\ttfamily{#1}}}

\newcommand{\beann}{ \begin{eqnarray*}}
\newcommand{\eeann}{ \end{eqnarray*}  }

\newcommand{\Lameff}{\Lambda_\mathrm{dec}}

\newcommand{\gbarnl}{\overline{g}_{\tl}}
\newcommand{\gbarnq}{\overline{g}_{\tq}}

\newcommand{\nq}{N_\mathrm{q}}
\newcommand{\tq}{\mathrm{q}}
\newcommand{\tl}{\mathrm{l}}
\newcommand{\Lamq}{\Lambda_{\tq}}
\newcommand{\Laml}{\Lambda_{\tl}}

\newcommand{\nl}{N_\mathrm{l}}
\newcommand{\msbar}{{\rm \overline{MS\kern-0.05em}\kern0.05em}}
\newcommand{\lag}[1]{{\mathcal{L}}_{\rm {#1}}}
\newcommand{\rmO}{{\rm O}}
\newcommand{\mhad}{m^\mathrm{had}}
\newcommand{\gbar}{\bar{g}}
\def\mstar{m_*}
\def\gstar{g_*}
\newcommand{\mbar}{\kern1pt\overline{\kern-1pt m\kern-1pt}\kern1pt}
\newcommand{\Mc}{M_{\rm c}}

\newcommand{\rmd}{{\rm d}}
\def\fm{{\rm fm}}
\def\Mref{M_{\rm ref}}
\newcommand{\rme}{{\rm e}}

\newcommand{\nf}{N_\mathrm{f}}

\title{Perturbative versus non-perturbative decoupling of heavy quarks}

\ShortTitle{Perturbative versus non-perturbative decoupling of heavy quarks}

\author{
   \includegraphics[width=0.2\linewidth]{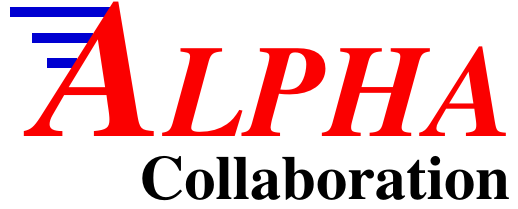}
}

\author{\speaker{Francesco Knechtli}\\
        Department of Physics, Bergische Universit{\"a}t Wuppertal\\
        Gaussstr.~20, 42119 Wuppertal, Germany\\
        E-mail: \email{knechtli@physik.uni-wuppertal.de}}

\author{Mattia Bruno\\
        Brookhaven Natl. Lab., Bldg. 510A\\
        20 Pennsylvania Street, Upton, NY 11973-5000, USA\\
        E-mail: \email{mbruno@quark.phy.bnl.gov}}

\author{Jacob Finkenrath\\
        CaSToRC, CyI Athalassa Campus \\
        20 Constantinou Kavafi Street, 2121 Nicosia, Cyprus\\
        E-mail: \email{j.finkenrath@cyi.ac.cy}}

\author{Bj{\"o}rn Leder\\
        Institutsrechenzentrum, Institut f{\"u}r Physik, Humboldt Universit{\"a}t zu Berlin\\
        Newtonstr.~15, 12489 Berlin, German\\
        E-mail: \email{leder@physik.hu-berlin.de}}

\author{Rainer Sommer\\
        John von Neumann Institute for Computing (NIC)\\
        DESY, Platanenallee~6, 15738 Zeuthen, Germany\\
        E-mail: \email{rainer.sommer@desy.de}}

\abstract{
We simulate a theory with $N_f=2$ heavy quarks of mass $M$. At energies much 
smaller than $M$ the heavy quarks decouple and the theory can be described 
by an effective theory which is a pure gauge theory to leading order in $1/M$. 
We present results for the mass dependence of ratios such as $t_0(M)/t_0(0)$. 
We compute these ratios from simulations and compare them to the perturbative 
prediction. The latter relies on a factorisation formula for the ratios which 
is valid to leading order in $1/M$.
\begin{flushright} WUB/15-07\\
      HU-EP-15/55\\
      DESY 15-219 \end{flushright}
}

\FullConference{The 33rd International Symposium on Lattice Field Theory\\
                 14 -18 July  2015\\
                 Kobe International Conference Center, Kobe, Japan}

\begin{document}

\section{Decoupling of heavy quarks}

At low energies $E\ll M$ heavy quarks of mass $M$ decouple. 
Their effects can then be described by an effective theory \cite{Weinberg:1978kz}.
The leading order of this effective theory is a theory where
the heavy quarks are removed.
Corrections to the leading order in the effective theory involve 
power corrections $(E/M)^n$ with $n\ge2$.
The heavy quarks leave traces through the renormalization of the gauge
coupling and the power corrections.
In this contribution we discuss
the renormalization effects. Power corrections are discussed in
\cite{Bruno:2014ufa,Knechtli:2014sta}.

We consider QCD with $\nq$ quarks and denote its Lambda parameter
by $\Lamq$. We will use the $\msbar$ scheme.
$\nl$ quarks are light and we set their mass to zero in the following.
$\nq-\nl$ quarks are heavy and their renormalization group invariant
(RGI) mass is $M$ (for its definition see \sect{s_matching}).
The Lagrangian of the effective theory $\lag{\rm dec}$ is defined
only in terms of $\nl$ light quarks and is given by a series in $1/M$
\begin{equation}\label{eq:Leff}
\lag{\rm dec} = \lag{QCD_\mathit{\nl}}
+ (1/M)^{2} \sum_i \omega_i \Phi_i +\rmO((\Lamq/M)^{4}) \,.
\end{equation} 
Here $\Phi_i$ are fields of dimension 6 and $\omega_i$ are dimensionless
parameters.
At leading order the effective theory is QCD with $\nl$ massless quarks.
QCD$_\mathit{\nl}$ has only one free parameter, namely the
gauge coupling which we denote by 
$\gbarnl(\mu/\Laml)$.
One can specify either a value for the coupling at some scale $\mu$ or 
equivalently the Lambda parameter.
Matching at leading order the effective theory with $\nl$ quarks to the full
theory with $\nq$ quarks (see \sect{s_matching}) yields
a relation $\Laml=\Lameff(M,\Lamq)$ which determines the Lambda parameter
of QCD$_\mathit{\nl}$ as a function of the heavy quark mass $M$ and the
Lambda parameter of QCD$_\mathit{\nq}$.

\section{Factorisation formula}

$\mhad$ denotes a hadron mass or a hadronic scale like 
$1/r_0$ \cite{pot:r0} or $1/\sqrt{t_0}$ \cite{flow:ML}.
The non-perturbative matching condition is
\begin{equation}\label{eq:npmatching}
\mhad_\tq = \mhad_\tl + \rmO((\Lamq/M)^2) \,,
\end{equation}
where $\mhad_\tq$ ($\mhad_\tl$) is the hadron mass computed in QCD$_\mathit{\nq}$
(QCD$_\mathit{\nl}$). \eq{eq:npmatching}
leads to the factorisation formula \cite{Bruno:2014ufa}
\begin{eqnarray}\label{eq:factorisation}
{\mhad_\tq(M) \over \mhad_\tq(0)}
& = &
    Q^{\mathrm{had}}_{\tl,\tq}
    \times P_{\tl,\tq}(M/\Lamq) + \rmO((\Lamq/M)^2) \,,
\end{eqnarray}
where
\begin{equation}\label{eq:P}
P_{\tl,\tq}(M/\Lamq) = \frac{\Laml}{\Lamq}
\end{equation}
and
\begin{equation}\label{eq:Q}
Q^{\mathrm{had}}_{\tl,\tq} = {\mhad_\tl/ \Laml \over \mhad_\tq(0)/\Lamq } \,.
\end{equation}
The factor $P_{\tl,\tq}$, \eq{eq:P}, can be computed in perturbation theory
and depends on $M$. It is universal in the sense that it does not 
depend on the hadronic scale.
The factor $Q^{\mathrm{had}}_{\tl,\tq}$ \eq{eq:Q} instead is non-perturbative and independent of $M$. 
It depends on the hadronic scale. The independence of $M$ relies on the 
observation that $\mhad_\tl/ \Laml$ is a pure number in QCD$_\mathit{\nl}$.
\eq{eq:factorisation} factorises the left hand side into a factor with a
perturbative expansion and a non-perturbative, but $M$-independent factor.

\section{Perturbative matching \label{s_matching}}

At leading order in $1/M$ matching imposes that observables computed
in QCD$_\mathit{\nq}$ are equal to observables computed in QCD$_\mathit{\nl}$.
In perturbation theory this leads to a relation between the $\msbar$-couplings 
$\gbarnl(\mu/\Laml)$ in QCD$_\mathit{\nl}$ and
$\gbarnq(\mu/\Lamq)$ in QCD$_\mathit{\nq}$:
\begin{equation}\label{eq:couplingmatch1}
  \gbarnl^2(\mu/\Laml)=\gbarnq^2(\mu/\Lamq) +\rmO(\gbarnq^4(\mu/\Lamq))\,.
\end{equation}
We choose the renormalization scale $\mu=\mstar$ \cite{Weinberg:1980wa,thresh:BeWe}
defined through the condition
\begin{equation}\label{eq:mstar}
\mbar(\mstar)=\mstar \,,
\end{equation}
where $\mbar(\mu)$ is the running quark mass.
The relation \eq{eq:couplingmatch1}
is known up to four loops \cite{Grozin:2011nk,Chetyrkin:2005ia}:
\begin{eqnarray}
   \gbarnl^2(\mstar/\Laml) &=&  \gbarnq^2(\mstar/\Lamq)\,
   C(\gbarnq(\mstar/\Lamq)) \nonumber 
   \\ && C(g)= 1+c_2g^4+c_3g^6+\ldots \quad (c_1=0) \label{eq:couplingmatch2}
\,.
\end{eqnarray}
The coefficients $c_2$ and $c_3$ can be found in \cite{Bruno:2014ufa}.
In the following we use the notation $\gstar=\gbarnq(\mstar/\Lamq)$.

From the matching relation \eq{eq:couplingmatch2} we can compute the
factor $P_{\tl,\tq}$ in \eq{eq:P}. The definition of the $\Lambda$ parameter 
in QCD with $\nf$ quarks and running coupling $\gbar$ is
\begin{equation}\label{eq:Lambda}
\Lambda = \mu \exp(I_g^\mathit{\nf}(\gbar(\mu))) \,,
\end{equation}
where $I_g^\mathit{\nf}(\gbar) = -\int^{\gbar} \rmd x\, \frac{1}{\beta_\mathit{\nf}(x)}$.
The $\beta_\mathit{\nf}$ function has the perturbative expansion
\begin{equation}\label{eq:beta}
 \beta_\mathit{\nf}(\bar g)  \buildrel {\bar g}\rightarrow0\over\sim 
 -{\bar g}^3 \left\{ b_0 + {\bar g}^{2}  b_1 + \ldots \right\}\,;\quad
 b_0(\nf)=\frac{1}{(4\pi)^2}\bigl(11-\frac{2}{3}\nf\bigr)\,,\;
 b_1(\nf)=\frac{1}{(4\pi)^4}\bigl(102-\frac{38}{3}\nf\bigr) \,.
\end{equation}
The precise definition of $I_g^\mathit{\nf}(\gbar)$ is given by
\begin{eqnarray}
\exp( I_g^\mathit{\nf}(\gbar)) &=& \left(b_0(\nf)\gbar^2\right)^{-b_1(\nf)/(2b_0(\nf)^2)} 
\rme^{-1/(2b_0(\nf)\gbar^2)} \nonumber \\
&& \times
           \exp \left\{-\int_0^{\gbar} \rmd x
          \left[\frac{1}{ \beta_\mathit{\nf}(x)}+\frac{1}{b_0(\nf)x^3}-\frac{b_1(\nf)}{b_0(\nf)^2x}
          \right]
          \right\} \,. \label{eq:Ig}
\end{eqnarray}
The factor $P_{\tl,\tq}$ in \eq{eq:P} is obtained
from \eq{eq:Lambda} and \eq{eq:couplingmatch2}:
\begin{equation}\label{eq:P2}
P_{\tl,\tq}(M/\Lamq) = 
\exp\left\{ I_g^\mathit{\nl}(\gstar\,\sqrt{C(\gstar)}) -  I_g^\mathit{\nq}(\gstar) \right\} \,,
\end{equation}
where $M$ is the RGI mass that corresponds to $\mstar$.

In order to evaluate \eq{eq:P2}
we need to determine the coupling $\gstar$.
The renormalization group invariant quark mass $M$ is defined from
the renormalized running mass $\mbar(\mu)$ by
\begin{equation}\label{eq:MRGI}
M = \mbar(\mu) \exp( I_m^\mathit{\nq}(\gbar)) \,, 
\end{equation}
where 
$I_m^\mathit{\nq}(\gbar)=-\int^{\gbar} \rmd x \,\frac{\tau_\mathit{\nq}(x)}{ \beta_\mathit{\nq}(x)}$.
The $\tau_\mathit{\nq}$ function has the perturbative expansion
\begin{equation}\label{eq:tau}
\tau_\mathit{\nq}(\bar g)  \buildrel {\bar g}\rightarrow0\over\sim 
 -{\bar g}^2 \left\{ d_0 + {\bar g}^{2}  d_1 + \ldots \right\}\,;\quad
d_0={8}/{(4\pi)^2} \,.
\end{equation}
The precise definition of $I_m^\mathit{\nq}(\gbar)$ is given by
\begin{equation}
\exp( I_m^\mathit{\nq}(\gbar)) = (2 b_0(\nq)\gbar^2)^{-d_0/(2b_0(\nq))}
   \exp \left\{- \int_0^{\gbar} \rmd x \left[{\tau_\mathit{\nq}(x) \over \beta_\mathit{\nq}(x)}
     - {d_0 \over b_0(\nq) x} \right] \right\} \,. \label{eq:Im}
\end{equation}
Using \eq{eq:Lambda} and \eq{eq:MRGI} we obtain ($\mu=\mstar$)
\begin{equation}\label{eq:MoL}
\frac{\Lamq}{M} =  \exp\left\{ I_g^\mathit{\nq}(\gstar)-I_m^\mathit{\nq}(\gstar)\right\} =
\exp \left\{-\int^{\gstar(M/\Lamq)} \rmd x\;
          \frac{1-\tau_\mathit{\nq}(x)}{ \beta_\mathit{\nq}(x)}    \right\} \,.
\end{equation}
Inverting this relation determines the coupling $\gstar$ as a function of
$M/\Lamq$ which in turn allows to compute $P_{\tl,\tq}$ through \eq{eq:P2}.
In the following we will use the notation $\Lambda\equiv\Lamq$.
\begin{figure}[t]\centering
  \resizebox{6.5cm}{!}{\includegraphics{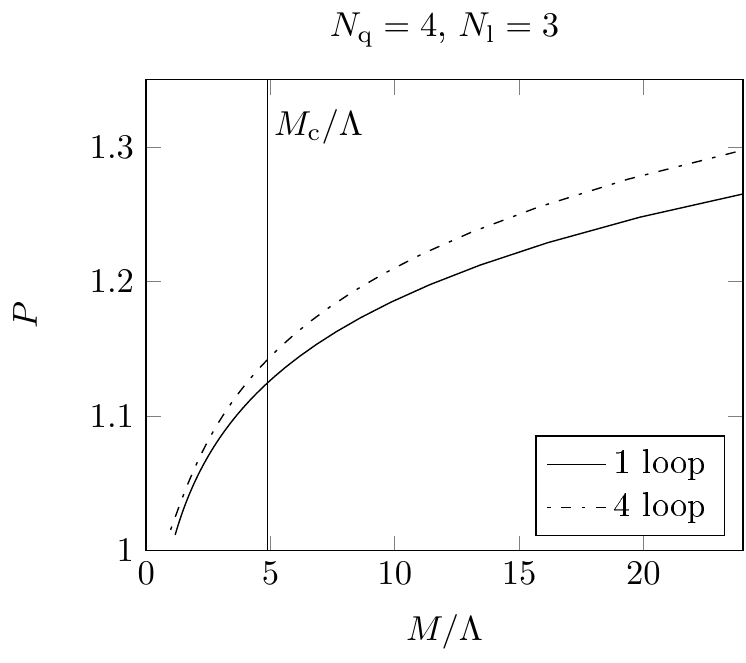}} \ \
  \resizebox{6.5cm}{!}{\includegraphics{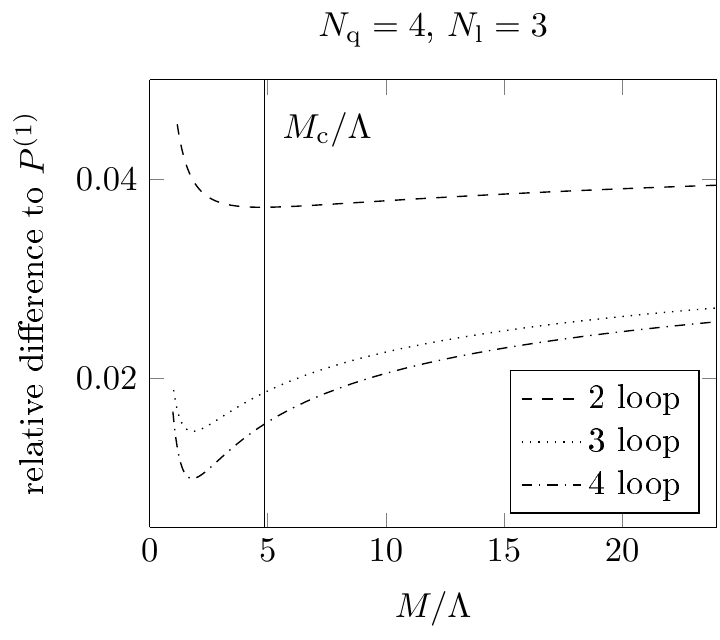}}
  \caption{Perturbative decoupling of the charm quark. 
The factor $P_{3,4}$ (left plot) and the corrections
$(P_{3,4}-P_{3,4}^{(1)})/P_{3,4}^{(1)}$ for different loop orders relative 
to the one-loop ``approximation'' $P_{3,4}^{(1)}=(M/\Lambda)^{2/27}$
(right plot).}
  \label{f:P34}
\end{figure}

For numerical applications we expand the integrands in \eq{eq:Ig} and \eq{eq:Im}
using \eq{eq:beta} and \eq{eq:tau}
and we evaluate the integrals numerically. 
For the $n$-loop expression we set $b_i=0$, $d_{i-1}=0$ for $i\ge n$.
A one-loop ``approximation'' is defined by setting $b_i=0\,,i>0$, $d_i=0$ :
\begin{equation}\label{eq:P1loop}
P_{\tl,\tq}^{(1)} = (M/\Lambda)^{\eta_0} \,,
\end{equation}
where $\eta_0 = 1 -{b_0(\nq) \over b_0(\nl)} > 0$.
In \fig{f:P34} we show the convergence of the perturbative 
expression for $P_{3,4}$, i.e., for the case of decoupling of the charm quark. 
The one-loop ``approximation'' is accidentally very close to the 
four-loop result.
In the right plots of \fig{f:P34} we show the relative correction 
$(P_{3,4}-P_{3,4}^{(1)})/P_{3,4}^{(1)}$ where $P_{3,4}$ is computed to 
$n$-loops ($n=2,3,4$).
The perturbative expansion appears to behave well
even for the case of the charm quark, where one would not 
have necessarily expected it to be so. 
More details on the calculation of $P_{\tl,\tq}$ 
will be presented in \cite{longpaper}.

\section{Non-perturbative results}

In order to check the factorisation formula \eq{eq:factorisation} and
the applicability of perturbation theory to compute the factor 
$P_{\tl,\tq}$ \eq{eq:P2},
we study a theory with $\nq=2$ heavy quarks and compare it to
Yang-Mills theory ($\nl=0$).
We simulate $\nq=2$ O($a$) improved Wilson quarks \cite{impr:csw_nf2}
with plaquette gauge action. The ensembles are listed in \tab{t:ens}.
The simulations with periodic boundary conditions (p) are done with the
MP-HMC algorithm \cite{lat10:marina} and those with open boundary conditions (o)
with the publicly available openQCD package \cite{algo:openQCD}.
We refer to \cite{Knechtli:2014sta} for further explanations.
\begin{table}[t]
 \centering
\begin{tabular}{cccccccc}
\toprule 
$\beta$ & $a$ [$\fm$]  & BC & $T\times L^3$   & $M/\Lambda_\msbar$ & kMDU & $\tau_{\rm exp}$ \\
\midrule
$5.3$ & $0.0658(10)$ & p & $64\times 32^3$   & 0.638(46) & 1.0 & 0.07 \\
      &              & p & $64\times 32^3$   & 1.308(95) & 2.0 & 0.05 \\
      &              & p & $64\times 32^3$   & 2.60(19)  & 2.0 & 0.04 \\
\midrule
$5.5$ & $0.0486($\phantom{0}$7)$  & o & $120\times 32^3$& 0.630(46) & 8.5 & 0.15 \\ 
      &              & o & $120\times 32^3$& 1.282(93) & 8.1 & 0.12 \\ 
      &              & p & $96\times 48^3$ & 2.45(18) & 4.0 & 0.10 \\
\midrule
$5.7$ & $0.0341($\phantom{0}$5)$  & o & $192\times 48^3$& 0.587(43) & 4.0  & 0.28 \\
      &              & o & $192\times 48^3$& 1.277(94) & 4.2 & 0.24 \\
      &              & o & $192\times 48^3$& 2.50(18) & 8.5 & 0.20 \\
\bottomrule
\end{tabular}
\caption{The decoupling ensembles.}
\label{t:ens}
\end{table}

We consider the scale $\mhad = 1/\sqrt{t_0}$ defined from the Wilson flow \cite{flow:ML}.
The factorisation formula for its mass-dependence reads, cf. \eq{eq:factorisation}
\begin{equation}\label{eq:t0factorisation}
\sqrt{t_0(M) / t_0(0)} = 1 / (P_{0,2}Q^{\sqrt{t_0}}_{0,2}) + \rmO((\Lambda/M)^{2})
\end{equation}
We compute from the simulations $t_0(M)/a^2$ at three values of the 
heavy quark mass close to the target values 
$M_\mathrm{t}/\Lambda=0.59$, $1.28$ and $2.50$. 
They correspond to approximately $\Mc/8$, $\Mc/4$ and $\Mc/2$ 
($\Mc$ is the charm quark mass).

The RGI mass $M$ and the ratio $M/\Lambda$ are computed as 
explained in \cite{Knechtli:2014sta}.
The data of the simulations in \tab{t:ens} are corrected for 
small mismatches compared to the target values $M_\mathrm{t}/\Lambda$. This
is done by fitting the $\beta=5.7$ data to the form
\begin{equation}\label{eq:massfit}
t_0(M)/a^2 = s_1\left(M/\Lambda\right)^\alpha \,.
\end{equation}
We get $\alpha=-0.246(5)$ which is close to $-2\eta_0=-0.242424$.
The corrected values $t_0(M_\mathrm{t})$ are computed as
\begin{equation}\label{eq:masscorr}
\ln(t_0(M_\mathrm{t})/a^2) = \ln(t_0(M)/a^2) + \alpha\ln(M_\mathrm{t}/M) \,.
\end{equation}
In order to keep the O($a$) improved coupling 
$\tilde{g}_{0}^{2}=(1+b_g(g_{0}^{2})\nq am) g_{0}^{2}$ fixed, we correct 
\begin{equation}\label{eq:bgcorr}
t_0(M_\mathrm{t})/a^2 \longrightarrow 
(1+2\times0.098 \nq am)\,t_0(M_\mathrm{t})/a^2 \,,
\end{equation}
where $m$ is the PCAC mass.

In order to compute the ratio in \eq{eq:t0factorisation} we need
the value $t_0(0)/a^2$ in the chiral limit. The latter is known only for 
$\beta=5.3$ and $\beta=5.5$ from \cite{Bruno:2013gha}. 
For our smallest lattice spacing $a(\beta=5.7)$ we use
\begin{equation}\label{eq:t0mref}
 \left.\sqrt{t_0(M) / t_0(0)}\right|_{a(5.7)} \approx 
 \left.\sqrt{t_0(M) / t_0(\Mref)}\right|_{a(5.7)} \times
 \lim_{a\to a(5.7)} \sqrt{t_0(\Mref) / t_0(0)} \,,
\end{equation}
where the reference mass is chosen to be our lightest mass 
$\Mref/\Lambda=0.59$. The limit in the second factor 
in \eq{eq:t0mref} is computed by a linear extrapolation of the data
$\sqrt{t_0(\Mref) / t_0(0)}$ at $\beta=5.3$ and $5.5$ as a function
of $a^2/(8t_0(\Mref))$.
We add to the error of the extrapolation half of the difference
between the $\beta=5.7$ and $\beta=5.5$ values.
This error is added linearly to the total error in \eq{eq:t0mref}.
\begin{figure}[t]\centering
  \resizebox{10cm}{!}{\includegraphics{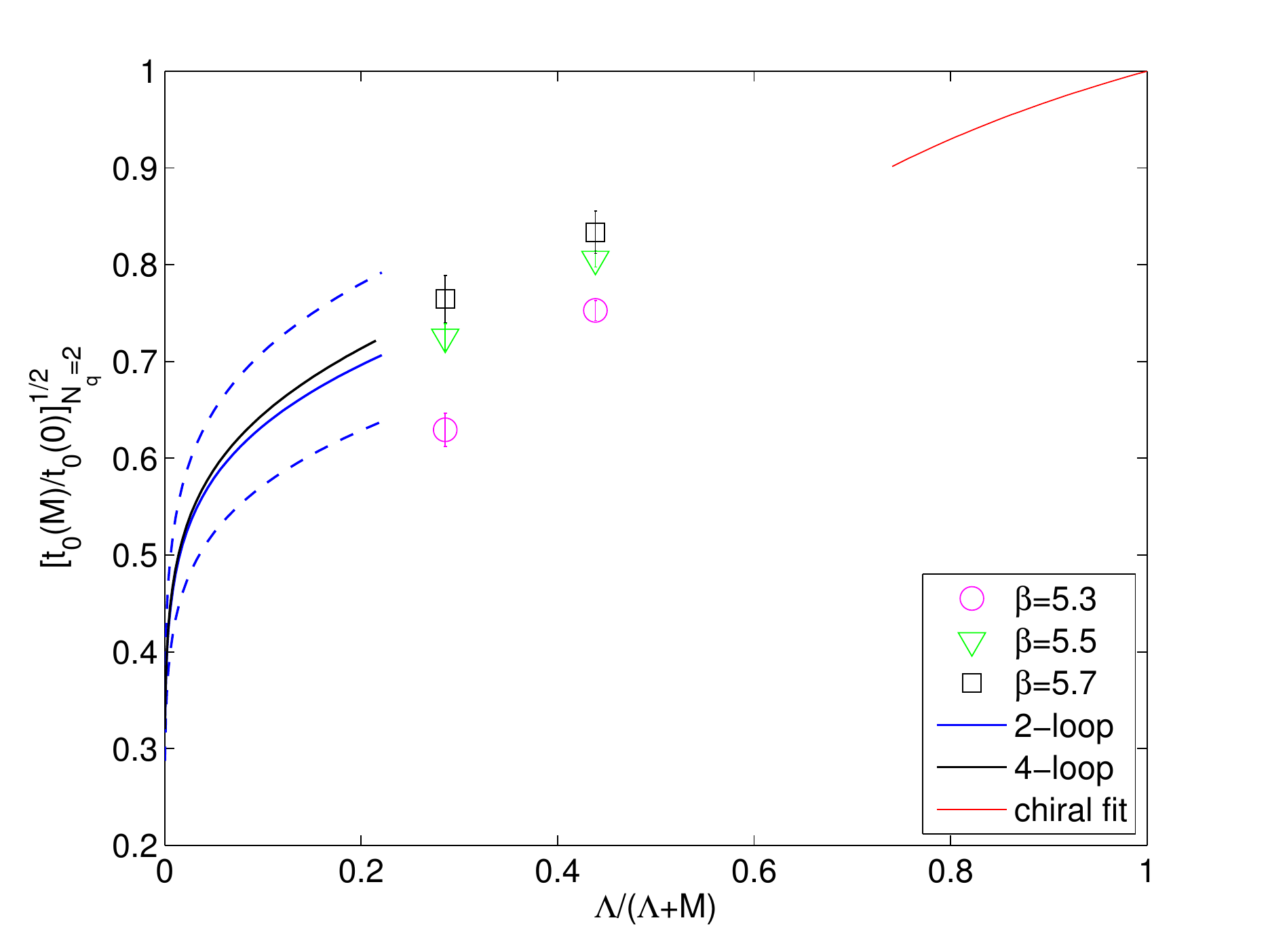}}
  \caption{The mass dependence of the ratio $\sqrt{t_0(M) / t_0(0)}$.}
  \label{f:bigplot}
\end{figure}

The results for the ratio $\sqrt{t_0(M) / t_0(0)}$ at
$M_\mathrm{t}/\Lambda=1.28$ and $2.50$ are shown by the 
symbols in \fig{f:bigplot}.
We compare them to the factorisation formula \eq{eq:t0factorisation},
where the factor $P_{0,2}$ is computed to 2- (blue line) and 4-loops 
(black line). The error on the factorization formula comes from 
$Q^{\sqrt{t_0}}_{0,2}=[\sqrt{t_0(0)}\Lambda]_{\nq=2} / [\sqrt{t_0}\Lambda]_{\nq=0}
\simeq 1.19(13)$ and is displayed by the dashed blue lines only for the 2-loop 
curve. The value of $Q^{\sqrt{t_0}}_{0,2}$ is obtained from
$Q^{r_0}_{0,2} = [\Lambda r_0(0)]_{\nq=2}/[\Lambda r_0]_{\nq=0} = 1.30(14)$
known from previous works \cite{mbar:pap1,alpha:lambdanf2}
and $[\sqrt{t_0(0)}/r_0(0)]_{\nq=2}/[\sqrt{t_0}/r_0]_{\nq=0} \simeq 0.915$ from
\cite{Sommer:2014mea}.
Within 10\% accuracy, the perturbative prediction for the mass dependence
of $\sqrt{t_0(M) / t_0(0)}$ agrees with our simulation results at $\beta=5.7$
for masses of about half the charm quark mass. 
For completeness, in \fig{f:bigplot} the red line to the right shows
the mass dependence in the chiral limit \cite{Sommer:2014mea,Bruno:2013gha}.

\section{Conclusions}
Perturbation theory seems to be reliable for decoupling of heavy quarks at 
leading order in $1/M$ even at the charm quark mass.
Our data from simulations of $\nq=2$ O($a$) improved Wilson quarks
shown in \fig{f:bigplot}
match the factorisation formula \eq{eq:factorisation} for the mass 
dependence of hadronic scales.
A careful continuum limit of the data in \fig{f:bigplot} will be addressed 
in the near future combined with twisted mass simulations.

{\bf Acknowledgements.}
We gratefully acknowledge the computer resources
granted by the Gauss Centre for Supercomputing (GCS) 
through the John von Neumann Institute for Computing (NIC) on the GCS share
of the supercomputer JUQUEEN at JSC,
with funding by the German Federal Ministry of Education and Research
(BMBF) and the German State Ministries for Research
of Baden-W\"urttemberg (MWK), Bayern (StMWFK) and
Nordrhein-Westfalen (MIWF).
We are further grateful for
computer time allocated for our project
on the Konrad and Gottfried computers at 
the North-German Supercomputing Alliance HLRN and
on the Cheops computer at the University of Cologne (financed
by the Deutsche Forschungsgemeinschaft).
This work is supported by the Deutsche Forschungsgemeinschaft
in the SFB/TR~09 and the SFB/TR~55.

\end{document}